\documentclass[11pt,a4paper]{article}
\usepackage{jheppub_kim}

\usepackage{pdflscape}
\usepackage{amsmath}
\usepackage{amssymb}
\usepackage{dcolumn}
\usepackage{bm}
\usepackage{color}
\usepackage{epsfig}
\usepackage{amsfonts}
\usepackage{graphicx}
\usepackage{subfigure}
\usepackage{dcolumn}

\newcommand{\be}{\begin{equation}}
\newcommand{\ee}{\end{equation}}
\newcommand{\bea}{\begin{eqnarray}}
\newcommand{\eea}{\end{eqnarray}}

\def\be{\begin{equation}}
\def\ee{\end{equation}}
\def\bea{\begin{eqnarray}}
\def\eea{\end{eqnarray}}

\begin{document}

\title{Interacting Holographic Extended Chaplygin Gas and Phantom Cosmology in the Light of BICEP2}
\author[a]{J. Sadeghi}
\author[a]{H. Farahani}
\author[b]{B. Pourhassan}

\affiliation[a]{Department of Physics, Faculty of Basic Sciences, University of Mazandaran, P. O. Box 47416-95447, Babolsar, IRAN}
\affiliation[b]{School of Physics, Damghan University, Damghan, Iran}

\emailAdd{pouriya@ipm.ir}
\emailAdd{h.farahani@umz.ac.ir}
\emailAdd{b.pourhassan@du.ac.ir}

\abstract{In this paper, we study the holographic dark energy
density and interacting extended Chaplygin gas energy density in the Einstein gravity.
We reconstruct the scalar field and the scalar potential describing the extended
Chaplygin gas. In the special case, we obtain energy density and investigate some cosmological parameters. Assuming interaction between components we find energy density for some different parametrization of total EoS. We analyze tensor to scalar ratio and use recent observational data of BICEP2 to fix the model parameters.}


\maketitle

\section{Introduction}
Recent discovery of the primordial gravitational waves by the BICEP2 experiment \cite{P1,P2} allows us to reconstruct some cosmological models, for example by calculation of the tensor-to-scalar ratio as $r=0.20^{+0.07}_{-0.05}$. An interesting cosmological model to describe universe is based on Chaplygin gas (CG) equation of state \cite{P3,P4},
\begin{equation}\label{s1}
p=-\frac{B}{\rho},
\end{equation}
where $B$ is a constant parameter. The CG was not consistent good with observational data, therefore, the generalized Chaplygin gas (GCG) introduced \cite{P5,P6,P7},
\begin{equation}\label{s2}
p=-\frac{B}{\rho^{\alpha}},
\end{equation}
where $\alpha$ is a constant parameter. The GCG can unify dark matter and dark energy. However, observational data ruled out such a proposal. Then, the modified Chaplygin gas (MCG) \cite{P8} proposed,
\begin{equation}\label{s3}
p=A\rho-\frac{B}{\rho^{\alpha}},
\end{equation}
where $A$ is a constant parameter. Recently, the generalized cosmic Chaplygin gas (GCCG) model also introduced \cite{P9,P10},
\begin{equation}\label{s4}
p=-\rho_{GCCG}^{-\alpha}\left[\frac{A}{1+\omega}-1+\left(\rho_{GCCG}^{1+\alpha}-\frac{A}{1+\omega}+1\right)^{-\omega}\right],
\end{equation}
in such a way that the resulting models can be made stable and free from unphysical behaviors even
when the vacuum fluid satisfies the phantom energy condition \cite{P11}. It is then straightforward to construct modified cosmic Chaplygin gas \cite{P12,P13,P14},
\begin{equation}\label{s5}
p=A\rho-\rho_{GCCG}^{-\alpha}\left[\frac{A}{1+\omega}-1+\left(\rho_{GCCG}^{1+\alpha}-\frac{A}{1+\omega}+1\right)^{-\omega}\right],
\end{equation}
It is also possible to consider viscosity in above models \cite{P151,P152,P153,P15,P16,P17,P18,P19,P20,P21,P22}.\\
Another way to study dark energy arises from holographic principle states that
the number of degrees of freedom related directly to the entropy scales with the enclosing
area of the system. In that case the total energy of the system with size $L$
should not exceed the mass of the same black hole size. It means that,
\begin{equation}\label{s6}
L^{3}\rho\leq L M_{p}^{2},
\end{equation}
where $\rho$ is the quantum zero-point energy density, also $M_{p}$ denotes Planck mass. Then, its holographic
energy density is given by the following expression \cite{P23,P24,P25,P26,P27},
\begin{equation}\label{s7}
\rho=\frac{3c^{2} M_{p}^{2}}{L^{2}},
\end{equation}
where $c$ is usually constant parameter, while there is possibility to consider non-constant $c$ \cite{P28}. The holographic model of dark energy based on Chaplygin gas equation of state already investigated for GCG, now we can construct new model based on the extended Chaplygin gas equation of state \cite{P29,P30,P31,P32},
\begin{equation}\label{s8}
p=\sum{A_{n}\rho^{n}}-\frac{B}{\rho^{\alpha}},
\end{equation}
where $n$ is a new parameter of the model. It is obvious that the $n=1$ with $A_{1}=A$ reduced to MCG.
\section{FRW cosmology}
The Friedmann-Robertson-Walker (FRW) Universe is described by the following metric,
\begin{equation}\label{s9}
ds^2=dt^2-a(t)^2(dr^2+r^{2}d\Omega^{2}),
\end{equation}
where $d\Omega^{2}=d\theta^{2}+\sin^{2}\theta d\phi^{2}$. Also $a(t)$
represents time-dependence scale factor. It yields to the following Friedmann equations,
\begin{equation}\label{s10}
H=\frac{\dot{a}}{a}=\frac{\rho}{3 M_{p}^{2}},
\end{equation}
and,
\begin{equation}\label{s11}
2\frac{\ddot{a}}{a}+(\frac{\dot{a}}{a})^{2}=-\frac{p}{M_{p}^{2}},
\end{equation}
where $H$ is Hubble expansion parameter and over dot denotes derivative with respect to cosmic time $t$. Also, one can obtain conservation equation as follow,
\begin{equation}\label{s12}
\dot{\rho}+3H(p+\rho)=0,
\end{equation}
where $p$ and $\rho$ are pressure and energy density obeying the relation (\ref{s8}). Here, we would like to consider the second order term which recovers quadratic barotropic equation of state,
\begin{equation}\label{s13}
p=A\rho+A_{2}\rho^{2}-\frac{B}{\rho^{\alpha}},
\end{equation}
where $\alpha$, $A_{1}$, $A_{2}$ and $B$ are free parameters of the model. In order to simplify calculations and reduce free parameters of the model we assume the following relations,
\begin{eqnarray}\label{s14}
\alpha&=&1,\nonumber\\
A&=&A_{2}-1,\nonumber\\
B&=&2A_{2}.
\end{eqnarray}
This is the simplest choice to have analytical expression for the energy density. Above choice relates $A$ and $B$ to new coefficient $A_{2}$ and we can study impact of additional term corresponding to the quadratic barotropic equation of state. Therefore, only free parameter of the model is $A_{2}$. Then, one can obtain,
\begin{equation}\label{s15}
\rho_{ECG}=1+\frac{2+\sqrt{-1+5e^{\frac{9}{2}}a^{30A_{2}}}}{-1+e^{\frac{9}{2}}a^{30A_{2}}},
\end{equation}
where we use the fact that $0\leq f\equiv \tan^{-1}(\rho+1)\leq1.5$, and set $f=\frac{3}{4}$. The equation (\ref{s15}) tells that the energy density have only physical meaning if $e^{\frac{9}{2}}a^{30A_{2}}>1$. It means that $a>e^{-\frac{3}{20A_{2}}}$, which is critical value of the scale factor at the early universe. In another word there is a minimum value for the scale factor $a=const\neq0$ at the early time cosmology. We can interpreted it as initial value of the scale factor after big bang. We should note that the energy density of (\ref{s15}) is corresponding to non-interacting case. We will find another expression in the case of interacting model.
\section{Interacting cosmology}
Now, we assume that there is an interaction between extended Chaplygin gas energy density $\rho$ and a cold
dark matter (CDM) with $\omega_{m}=0$. The conservation equation now separated as follows,
\begin{equation}\label{s16}
\dot{\rho}_{ECG}+3H(
1+\omega_{ECG})\rho_{ECG}=-Q,
\end{equation}
and,
\begin{equation}\label{s17}
\dot{\rho}_{m}+3H\rho_{m}=Q,
\end{equation}
where we assumed $\rho=\rho_{ECG}+\rho_{m}$, with CDM energy density $\rho_{m}$. Moreover, using the equation (\ref{s13}) we have,
\begin{equation}\label{s18}
\omega_{ECG}=\frac{p_{ECG}}{\rho_{ECG}}
=A_{2}-1+A_{2}\rho_{ECG}-\frac{2A_{2}}{\rho_{ECG}^{2}}.
\end{equation}
The interaction term assumed as follow,
\begin{equation}\label{s19}
Q=b\rho_{ECG},
\end{equation}
where $b$ is an interaction coefficient which can interpreted as decaying of the extended Chaplygin gas component into CDM. It may considered as constant of varying quantity.\\
Equations (\ref{s16}), (\ref{s17}) and (\ref{s19}) may be combine as follow,
\begin{equation}\label{s20}
\dot{\varrho}=3H\varrho\omega_{ECG}+(1+\varrho)b,
\end{equation}
where,
\begin{equation}\label{s21}
\varrho\equiv\frac{\rho_{m}}{\rho_{ECG}}.
\end{equation}
It help us to find simple expression for $\varrho$ under assumption of $H\propto\omega_{ECG}^{-1}$. This assumption may reasonable because according to the Friedmann equation, $H\propto\rho$ and $\rho\propto\omega^{-1}$. Also, negative nature of $\omega_{ECG}$ suggests to choose $H=-h\omega_{ECG}^{-1}$ where $h$ is an arbitrary constant. So, we have an inhomogeneous first order linear differential equation with the following solution,
\begin{equation}\label{so21}
\varrho=-\frac{b}{b-3h}\left(1-e^{(b-3h)t}\right)+\varrho_{0}e^{(b-3h)t},
\end{equation}
where $\varrho_{0}$ is value of $\varrho$ at $t=0$ (early universe). It is obvious that $h>b/3$ is necessary condition to have agreement with observational data. However this is only special solution which may be useful for special goals. At the early universe, which was matter dominant, value of $\varrho=\varrho_{0}$ is large which means that $\rho_{m}\gg\rho_{ECG}$.\\
On the other hand at the late time, which is energy dominant ($\rho_{ECG}\gg\rho_{m}$), and we have $\varrho=\frac{b}{3h-b}\ll1$, which means $h\gg2b/3$.\\
In the general case, we try to rewrite the equations (\ref{s16}) and (\ref{s17}) as follows,
\begin{equation}\label{s22}
\dot{\rho}_{ECG}+3H(1+\omega_{ECG}^{eff})\rho_{ECG}=0,
\end{equation}
and,
\begin{equation}\label{s23}
\dot{\rho}_{m}+3H(1+\omega_{m}^{eff})\rho_{m}=0,
\end{equation}
where we defined,
\begin{eqnarray}\label{s24}
\omega_{ECG}^{eff}&\equiv&\omega_{ECG}+\frac{b}{3H},\nonumber\\
\omega_{m}^{eff}&\equiv&-\frac{1}{\varrho}\frac{b}{3H}.
\end{eqnarray}
By using the following definitions,
\begin{eqnarray}\label{s25}
\Omega_{ECG}&\equiv&\frac{\rho_{ECG}}{3H^{2}M_{p}^{2}},\nonumber\\
\Omega_{m}&\equiv&\frac{\rho_{m}}{3H^{2}M_{p}^{2}},
\end{eqnarray}
one can obtain,
\begin{equation}\label{s26}
\Omega_{m}+\Omega_{ECG}=1,
\end{equation}
which yields to the following relation,
\begin{equation}\label{s27}
\varrho=\frac{1-\Omega_{ECG}}{\Omega_{ECG}}.
\end{equation}
\begin{figure}[h!]
 \begin{center}$
 \begin{array}{cccc}
\includegraphics[width=70 mm]{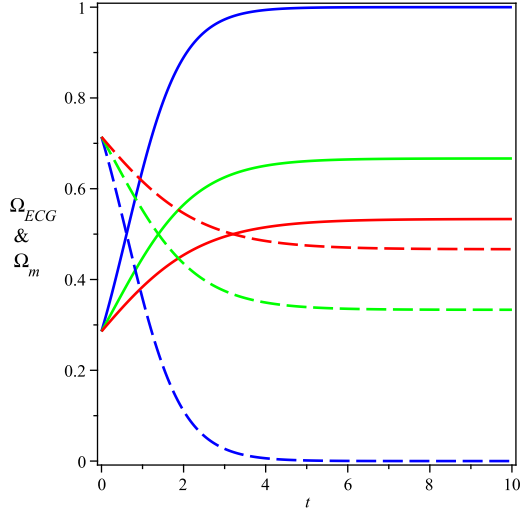}
 \end{array}$
 \end{center}
\caption{$\Omega_{m}$ (dash lines) and $\Omega_{ECG}$ (solid lines) in terms of time for $\varrho=2.5$ and $h=0.5$. $b=0$ (blue), $b=0.5$ (green), $b=0.7$ (red)}
\label{fig:1}
\end{figure}
So, we can use the relation (\ref{so21}) to obtain time-dependent $\Omega_{ECG}$ and therefore $\Omega_{m}$ using the relation (\ref{s26}).\\
In the Fig. \ref{fig:1} we can see that presence of interaction ($b\neq0$) is necessary to have $\Omega_{ECG}$ and $\Omega_{m}$ in the same order (at the late time) which is a possible solution of cosmic coincidence problem \cite{copeland-2006,Nobbenhuis}.\\
Then, the effective equation of state reads as,
\begin{equation}\label{s28}
\omega_{ECG}^{eff}=A_{2}-1+A_{2}\rho_{ECG}-\frac{2A_{2}}{\rho_{ECG}^{2}}+\frac{b}{3H}.
\end{equation}
As we told already, it is possible to choose varying interaction coefficient $b$. It may depend on densities, scale factor, Hubble parameter or explicit time.
Here we use the following relation of the interaction coefficient,
\begin{equation}\label{s29}
b=3g(1+\varrho)H=\frac{3gH}{\Omega_{ECG}},
\end{equation}
where $g$ is coupling constant. It help us to simplify equations. Therefore, one can obtain,
\begin{equation}\label{s30}
\omega_{ECG}^{eff}=A_{2}-1+3A_{2}H^{2}M_{p}^{2}\Omega_{ECG}-\frac{2A_{2}}{9H^{4}M_{p}^{4}\Omega_{ECG}^{2}}+\frac{g}{\Omega_{ECG}}.
\end{equation}
Using the holographic dark energy density of the extended Chaplygin gas,
\begin{equation}\label{s31}
\rho_{ECG}=\frac{3c^{2} M_{p}^{2}}{L^{2}},
\end{equation}
we can rewrite the equation (\ref{s30}) as follow,
\begin{equation}\label{s32}
\omega_{ECG}^{eff}=A_{2}-1+\frac{3A_{2}M_{p}^{2}c^{2}}{L^{2}}-\frac{2A_{2}L^{4}}{9c^{4}M_{p}^{4}}+\frac{gL^{2}H^{2}}{c^{2}}.
\end{equation}
Since Hubble parameter is depend on scale factor, and scale factor is depend on time, so easily we can rewrite the above equation as follow,
\begin{equation}\label{s33}
\omega_{ECG}^{eff}=\omega_{0}+\omega_{1}f(t),
\end{equation}
where $\omega_{0}$ and $\omega_{1}$ are constants defined by,
\begin{eqnarray}\label{s34}
\omega_{0}&\equiv&-1+A_{2}\left(1+\frac{3M_{p}^{2}c^{2}}{L^{2}}-\frac{2L^{4}}{9c^{4}M_{p}^{4}}\right),\nonumber\\
\omega_{1}&\equiv&\frac{gL^{2}}{c^{2}},
\end{eqnarray}
where $\omega_{0}=-1$ and $\omega_{1}=0$ are corresponding to current value of the EoS parameter, and also comparing the equation (\ref{s32}) with (\ref{s33}) using definitions (\ref{s34}) gives,
\begin{equation}\label{s35}
f(t)\equiv H^{2}.
\end{equation}
Assuming $\omega_{ECG}^{eff}\approx\omega_{tot}$ (the case of negligible $\omega_{m}^{eff}$) allows us to use some well-known parametrization of EoS \cite{P33} to find time-dependence scale factor. These parametrization are based on redshift $z$ related to the scale factor via,
\begin{equation}\label{s36}
a=\frac{1}{1+z}.
\end{equation}
We will study four different forms of parametrization.
\subsection{redshift parametrization}
One of the simplest parametrization given by \cite{P34},
\begin{equation}\label{s37}
\omega_{tot}=\omega_{0}+\omega_{1}(\frac{1-a}{a})
\end{equation}
Comparing the equations (\ref{s33}), (\ref{s35}) and (\ref{s37}) tells that,
\begin{equation}\label{s37.5}
f(t)=H^{2}=\frac{1-a}{a},
\end{equation}
so we can obtain scale factor as follow,
\begin{equation}\label{s38}
a=\frac{1+\sin{t}}{2}.
\end{equation}
It is quit strange result, as we expected previously, the initial value of the scale factor is not zero. The scale factor begin with finite value and grow through expansion of universe. At the some time, expansion finished and scale factor decreases to zero, which is end of the world. Again the universe begin expanding from $a=0$ to a maximum value, and so on. It is completely characteristic of sin functions.\\
We can obtain the extended Chaplygin gas energy density at the early or late universe as follow,
\begin{equation}\label{s39}
\rho_{ECG}=\frac{(b+3A_{2})e^{-(b+3A_{2})t}}{C(b+3A_{2})-3A_{2}e^{-(b+3A_{2})t}},
\end{equation}
where $C$ is an integration constant and $b=3g$ because at the early time Hubble expansion has large value while $\Omega_{ECG}$ has infinitesimal value so one can assume $H\approx\Omega_{ECG}^{-1}$, on the other hand at the late time, both $H$ and $\Omega_{ECG}$ yield to a constant so one can assume $H/\Omega_{ECG}\propto const$. So, we can obtain interaction term $Q$ and using the equation (\ref{s17}) we can obtain matter energy density as follow,
\begin{equation}\label{s40}
\rho_{m}={\frac { \left( -3A_{2}-{b} \right) \cos
 \left( 3\,t \right) + \left( 9\,{b}+27A_{2} \right) \sin \left( 2
\,t \right) + \left( 45\,{b}+135A_{2} \right) \cos \left( t
 \right) -30\,{b}t+36\,{\it}\,A_{2}-90\,tA_{2}}{ 9b^{-1}\left( 10+15\,\sin
 \left( t \right) -6\,\cos \left( 2\,t \right) -\sin \left( 3\,t
 \right)  \right) A_{2}}},
\end{equation}
where we assumed $C$ as infinitesimal constant, and set new integration constant equal $b$. Comparing the equation (\ref{s39}) with non-interacting version (\ref{s15}) is interesting. If we switch of interaction in the equation (\ref{s39}), then we have,
\begin{equation}\label{s41}
\rho_{ECG}(b=0)=\frac{e^{-3A_{2}t}}{C-e^{-3A_{2}t}},
\end{equation}
We expect an equivalent between relations (\ref{s41}) and (\ref{s15}). It is possible if we choose $a\propto e^{-c\frac{t}{5}-\frac{3}{20A_{2}}}$, where $c$ is an arbitrary constant. So both equations (\ref{s15}) and (\ref{s39}) are coincide at the early universe. It is clear that the scale factor (\ref{s38}) and above exponential form may coincide at the early universe with infinitesimal $t$.
\subsection{scale factor parametrization}
It is also possible to choose following parametrization,
\begin{equation}\label{s44}
\omega_{tot}=\omega_{0}+\omega_{1}(1-a),
\end{equation}
known as scale factor parametrization \cite{P35}. Comparing the equations (\ref{s33}), (\ref{s35}) and (\ref{s44}) tells that,
\begin{equation}\label{s45}
f(t)=H^{2}=1-a,
\end{equation}
so we can obtain scale factor as follow,
\begin{equation}\label{s46}
a=1-\tanh^{2}{\frac{t}{2}}.
\end{equation}
\begin{figure}[h!]
 \begin{center}$
 \begin{array}{cccc}
\includegraphics[width=75 mm]{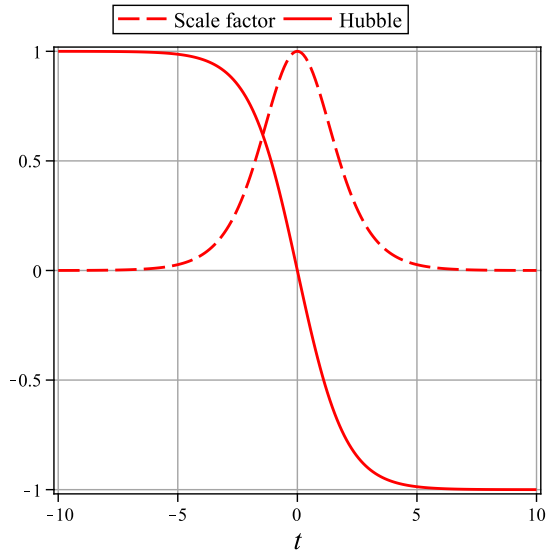}
 \end{array}$
 \end{center}
\caption{Scale factor (dashed line), and Hubble expansion parameter (solid line) in terms of time for the scale factor parametrization.}
\label{fig:2}
\end{figure}
It is quite unexpected result, the scale factor begin from finite initial value and decreases to zero, so there is no expanding universe if we set $t=0$ as initial time. In order to have agreement with observational data we should set $t=0$ corresponding to end of expansion. In that case the early universe begin at $t=-\infty$ expanding to the current stage, and continue expanding to $t=0$. After finishing expansion, the scale factor decreases to zero at $t=\infty$, which is end of the world. In this case the Hubble parameter has positive value at the early universe which yields to a constant at present. Then it decreases to zero at the end of expansion and take negative value to the end. In the Fig. \ref{fig:2} we can see behavior of scale factor and Hubble expansion parameter. We can say that current stage where Hubble parameter is a constant is within $t<-5$. So, at the early and late time, the Hubble expansion parameter behave as a constant and we can obtain similar result with previous subsection.
\subsection{logarithmic parametrization}
Another possible parametrization is given by \cite{P36},
\begin{equation}\label{s47}
\omega_{tot}=\omega_{0}+\omega_{1}\ln{\frac{1}{a}}
\end{equation}
Comparing the equations (\ref{s33}), (\ref{s35}) and (\ref{s47}) tells that,
\begin{equation}\label{s48}
f(t)=H^{2}=\frac{1}{a},
\end{equation}
so, we can obtain scale factor as follow,
\begin{equation}\label{s49}
a=\frac{t^{2}}{4},
\end{equation}
which yield to a simple relation for the Hubble parameter as,
\begin{equation}\label{s50}
H=\frac{2}{t}.
\end{equation}
Above solution tell that universe begin from $a=0$ at $t=0$ and expand to the end, so the Hubble parameter is decreasing function of time. We can obtain energy densities of the late and early time separately.\\
At the early time we can obtain the ECG density in terms of combination of Bessel functions of the first and second kinds respectively denoted by $BesselJ$ and $BesselY$,
\begin{equation}\label{s51}
\rho_{ECG}=\frac{2M_{p}^{2}\sqrt{3g}\left[BesselY(1-3A_{2},\frac{12M_{p}^{2}\sqrt{3A_{2}g}}{t})+BesselJ(1-3A_{2},\frac{12M_{p}^{2}\sqrt{3A_{2}g}}{t})\right]}
{t\sqrt{A_{2}}\left[BesselY(-3A_{2},\frac{12M_{p}^{2}\sqrt{3A_{2}g}}{t})+BesselJ(-3A_{2},\frac{12M_{p}^{2}\sqrt{3A_{2}g}}{t})\right]}.
\end{equation}
In that case from the equation (\ref{s17}) one can obtain,
\begin{equation}\label{s52}
\rho_{m}=\frac{18gM_{p}^{2}}{t^{2}}+\frac{C}{t^{6}},
\end{equation}
where $C$ is an arbitrary constant.\\
On the other hand at the late time we analyze two different cases of $g\gg1$ and $g\ll1$.\\
If coupling be very large ($g\gg1$), then we can obtain,
\begin{equation}\label{s53}
\rho_{ECG}\approx\frac{36gM_{p}^{2}}{t^{2}},
\end{equation}
which yields to the $\rho_{m}$ similar to the equation (\ref{s52}).\\
Finally for the case of weak coupling ($g\ll1$) we can obtain,
\begin{equation}\label{s54}
\rho_{ECG}\approx\sqrt{2+\frac{C}{t^{12A_{2}}}},
\end{equation}
where $C$ is an integration constant, while, interestingly, the matter density obtained as before which is given by the equation (\ref{s52}).
\section{Interacting extended Chaplygin gas and holographic phantom}
In order to complete our study we assume ECG interacting with Phantom field.
Phantom energy density and pressure are given by,
\begin{equation}\label{s55}
\rho_{ECG}=-\frac{1}{2}\dot{\phi}^{2}+V(\phi),
\end{equation}
and,
\begin{equation}\label{s56}
p_{ECG}=-\frac{1}{2}\dot{\phi}^{2}-V(\phi),
\end{equation}
which yield to the scalar potential and kinetic energy term as follow,
\begin{equation}\label{s57}
V(\phi)=\frac{1}{2}(1-\omega_{ECG})\rho_{ECG},
\end{equation}
and,
\begin{equation}\label{s58}
\dot{\phi}^{2}=-(1+\omega_{ECG})\rho_{ECG}.
\end{equation}
So, we can obtain,
\begin{equation}\label{s59}
V=\frac{1}{2}\left(2-2A_{2}-\frac{2+\sqrt{-1+5e^{\frac{9}{2}}a^{30A_{2}}}}{-1+e^{\frac{9}{2}}a^{30A_{2}}}
+\frac{2A_{2}}{\left[1+\frac{2+\sqrt{-1+5e^{\frac{9}{2}}a^{30A_{2}}}}{-1+e^{\frac{9}{2}}a^{30A_{2}}}\right]^{2}}\right)\frac{3c^{2}M_{p}^{2}}{2L^{2}},
\end{equation}
and,
\begin{equation}\label{s60}
\dot{\phi}^{2}=\left(-2A_{2}-\frac{2+\sqrt{-1+5e^{\frac{9}{2}}a^{30A_{2}}}}{-1+e^{\frac{9}{2}}a^{30A_{2}}}
+\frac{2A_{2}}{\left[1+\frac{2+\sqrt{-1+5e^{\frac{9}{2}}a^{30A_{2}}}}{-1+e^{\frac{9}{2}}a^{30A_{2}}}\right]^{2}}\right)\frac{3c^{2}M_{p}^{2}}{L^{2}},
\end{equation}
We can see that the scalar potential is decreasing function of the scalar field and may vanish at the late time. So, using the condition $V=0$ we can obtain late time value of the scale factor. In order to have real potential at the late time we should choose positive $A_{2}$. Then, using the $a\gg1$ approximation one can obtain,
\begin{equation}\label{s61}
a\approx \frac{M_{p}L\sqrt{3}}{c(\phi-\phi_{0})}.
\end{equation}
It is clear that increasing scalar field decreases value of the scale factor. It means that the scalar field has large value at the early universe while has small value at the late time. Then, using the relation,
\begin{equation}\label{s62}
H=\frac{\dot{\phi}}{a}\frac{da}{d\phi},
\end{equation}
we can obtain Hubble expansion parameter in terms of the scalar field. In that case we can obtain initial value of the scalar field as,
\begin{equation}\label{s63}
\phi(0)=\phi_{0}+\frac{\sqrt{3}M_{p}L}{c}e^{\frac{3}{20A_{2}}},
\end{equation}
\section{The tensor to scalar ratio}
\begin{figure}[h!]
 \begin{center}$
 \begin{array}{cccc}
\includegraphics[width=75 mm]{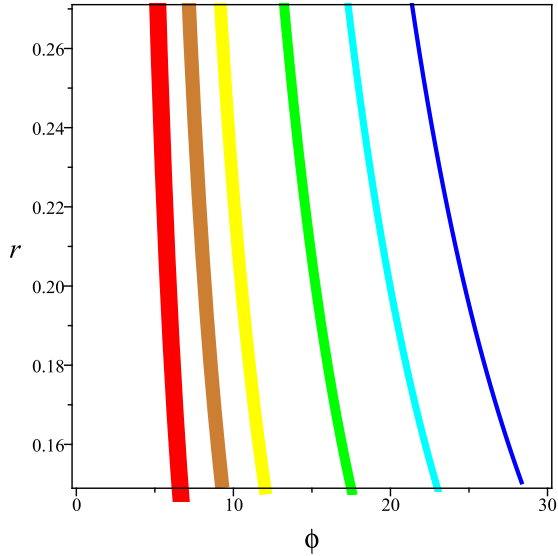}
 \end{array}$
 \end{center}
\caption{The tensor to scalar ratio in terms of the scalar field for $c=1$ and $\theta=0.3$. Left to right curves respectively corresponding to $A_{2}=-0.05, -0.075, -0.15, -0.2, -0.25$.}
\label{fig:3}
\end{figure}

We can use observational data to fix free parameters of the model. We use one of the recent and important results.
The tensor to scalar ratio $r$ is given by,
\begin{equation}\label{s64}
r\simeq16\epsilon_{H},
\end{equation}
where $\epsilon_{H}$ is one of the Hubble slow-roll parameters given by,
\begin{equation}\label{s65}
\epsilon_{H}=2\left(\frac{\frac{\partial H}{\partial\phi}}{H}\right)^{2}.
\end{equation}
We know that the current bound on $r$ by BICEP2 is $0.15\leq r\leq0.27$. So, we can fix free parameters of the model to gain agreement with observational data. In the Fig. \ref{fig:3} we can see that the tensor to scalar ratio is decreasing function of the scalar field. There are some regions where the tensor to scalar ratio is agree with recent BICEP2 data.\\
It is illustrated that decreasing $A_{2}$ needs increasing scalar field to have the tensor to scalar ratio in the expected range. From the equation (\ref{s61}) we can see that the scalar field and scale factor have inverse relation at the late time where value of the scale factor is high, therefore value of the scalar field is small. It means that the value of the parameter $A_{2}$ should be infinitesimally negative at current stage. It is completely agree with the nature of the second order terms which is important at the early universe. At the late time the lost term of the equation of state dominant.\\
\section{Conclusion}
In this paper, we give more investigation about recently proposed extended Chaplygin gas model in the light of BICEP2. Indeed, we construct holographic version of the extended Chaplygin gas with possibility of interaction with matter. We obtained scalar field and scalar potentials in terms of scale factor which yields to calculation of the tensor to scalar ratio. We considered second order term of the extended model which recovers quadratic barotropic equation of state. Under some assumption we reduced free parameters of the model to one. Numerically, we find that the model is in agreement with observational data. We used three well-known parametrization to obtain solution and discussed about the early and late time cosmology.\\
It is also possible to consider another parametrization such as,
\begin{equation}\label{Co1}
\omega_{tot}=\omega_{0}+\omega_{1}\ln{(2-a)},
\end{equation}
which introduced by the Ref. \cite{P33}.\\
It is also interesting to consider effect of viscosity in the model. These may part subjects of our future works.

\end{document}